\documentclass[
reprint,
 amsmath,amssymb,
 aps,
prb,
floatfix,
]{revtex4-2}

\usepackage{graphicx}%
\usepackage{dcolumn}%
\usepackage{bm}%
\usepackage[position=t,singlelinecheck=off,justification=raggedright,caption=false]{subfig}
\usepackage[]{natbib}
\usepackage[colorlinks,linkcolor=blue, urlcolor=blue,citecolor=blue]{hyperref}
\usepackage{multirow}

\begin{document}

\title{Spin-valley-layer coupling with dual control via stacking and electric field in antiferromagnetic bilayer Janus YIBr}
\author{Bo-Wen Yu}\email{yubowen@jnxy.edu.cn}
\affiliation{School of Physics and Electronic Engineering, Jining University, Qufu 273155,China}

\date{\today}

\begin{abstract}

The modification and enhancement of antiferromagnetic two-dimensional semiconductor is considered crucial for realizing novel electronic properties and facilitating promising applications. For this purpose, we investigate six antiferromagnetic 2D bilayer Janus YIBr structures with different stacking variations by means of first-principles calculation and an effective low-energy model. The calculation of magnetic anisotropy energy shows that the direction of easy axis varies with different stacking. First-principles-calculated energy bands reveal that there is a Dirac relativistic dispersion relation exists in the valence band in a wide energy window of 0.3 eV at least. The calculations for spin, atom properties and Berry curvature description show that there is spin, valley and layer coupling with spin splitting, valley polarization and valley Hall and layer Hall effects can be achieved in the bilayer Janus structures. Further analyses of the effect of external electric field can be used to control spin, valley and layer of the hole near the Fermi level and to realize an anomalous valley Hall effect. These can be useful in future exploration for novel properties, control methods and more functionalities in bilayer Janus structures.

\end{abstract}

\maketitle

\section{INTRODUCTION}

Two-dimensional materials have been extensively explored for their important phenomena and practical applications since the advent of graphene in 2004 \cite{G00,G01,G02}. Transition Metal Halides (TMHs) are famous for their unique properties in scientific research and engineering applications in the field of novel electronic devices. It is known that the asymmetric Janus structure of 2D TMHs can lead to richer properties and make it easier to control the properties of the materials. Actually, there have been many experimental and theoretical advances in Janus materials, driven by the rapid development of both experimental techniques and computational methods \cite{VCl2,YIBr,Rashba2,R01,J02,StackM,J05}.

It is known that spin splitting in 2D materials with zero net magnetization is a rapidly growing research direction\cite{Spinlayer,AFM1,AFM_MnBr,AFM3,AFM4,MA2D,MnPS3,HAFM,AFM5}. The SOC effect may couple the freedom of the spin and valley to create various properties and topological structures. Interlayer van der Waals interactions in a bilayer structure can also give rise to a richer array of properties, which can even introduce SOC from one layer to the other\cite{AFM2,J01,YU,J04}. The SOC can affect the valley polarization, response to an external field and carrier properties and electronic functionalities in semiconductor technology\cite{FV_2016,FV_L2023,QVHE1,mf1,AVHE1,AVHE2}.Interlayer coupling in bilayer Janus structures introduces a new dimension for tuning spin, valley and layer degrees of freedom\cite{LHE,FV_L2024,AFM4,YFI}. Due to the breaking of symmetry, Janus structures generate an intrinsic electric field and Rashba effect, making them ideal platforms for spin-valley-layer coupling\cite{VS2,YI2,NiI2,Rashba3,Janus1,Rashba1,J03,J06}. Considering recent experimental advances, it is highly desirable to explore new phenomena and novel effects induced in the antiferromagnetic 2D Janus structures.

Here, we investigate the bilayer Janus structures with different stacking types by means of first-principles methods and effective low-energy models. The magnetic anisotropy energy (MAE) shows that the direction of the easy axis is different with different stacking. The calculation of the band structure shows that a Dirac relativistic dispersion relation in the valence band in a wide energy window of 0.3 eV at least. Further analyses of spin and atom projection indicate that there are spin, valley and layer coupling with spin splitting and valley polarization in the valence band.The properties of edge of valence band can be indirectly controlled by altering the stacking configuration, thereby influencing the direction of the magnetic easy axis. The calculation of the Berry curvature description shows that valley Hall effect appears in the AA stackings and a layer Hall effect appears in the AB stackings. We also find that the external electric field can induce and enhance the spin splitting and realize an anomalous valley Hall effect, thereby controlling the properties of the holes. The detailed data and further analyses will be presented in the following.

\section{METHODOLOGY}

The first-principles calculations are performed with the projector-augmented wave (PAW) method within the density functional theory\cite{PAW}, implemented in the Vienna Ab-initio simulation package software (VASP) \cite{VASP}. The generalized gradient approximation (GGA)  by Perdew, Burke, and Ernzerhof (PBE)\cite{pbes} is used as the exchange-correlation functional. The  self-consistent calculations are carried out with a $\Gamma$-centered ($15\times 15\times 1$) Monkhorst-Pack grid\cite{MP}. The kinetic energy cutoff of the plane wave is set to 450 eV. The convergence criteria of the total energy and force are set to 10$^{-6}$ eV and 0.01 eV/\AA{}. The spin-orbit coupling (SOC) is applied in the calculation of band structure and lattice structure. The inter-layer vacuum thickness is set to at least 30 \AA{}. The dipole correction \cite{DP2,DP1} is applied in the calculation for the Janus structure. The Hubbard-U term (U = 2 eV) for Y is considered in the DFT+U scheme to improve energy band description. Dispersion corrections are taken into account via the Grimme approximation (DFT-D3)\cite{DFTD3}. The Berry curvature is calculated using the VASPBERRY code\cite{vaspberry}.

\section{RESULT AND DISCUSSION}

\subsection{Lattice structures and magnetic properties }

\begin{figure}[htbp]%
    \includegraphics[width=\columnwidth]{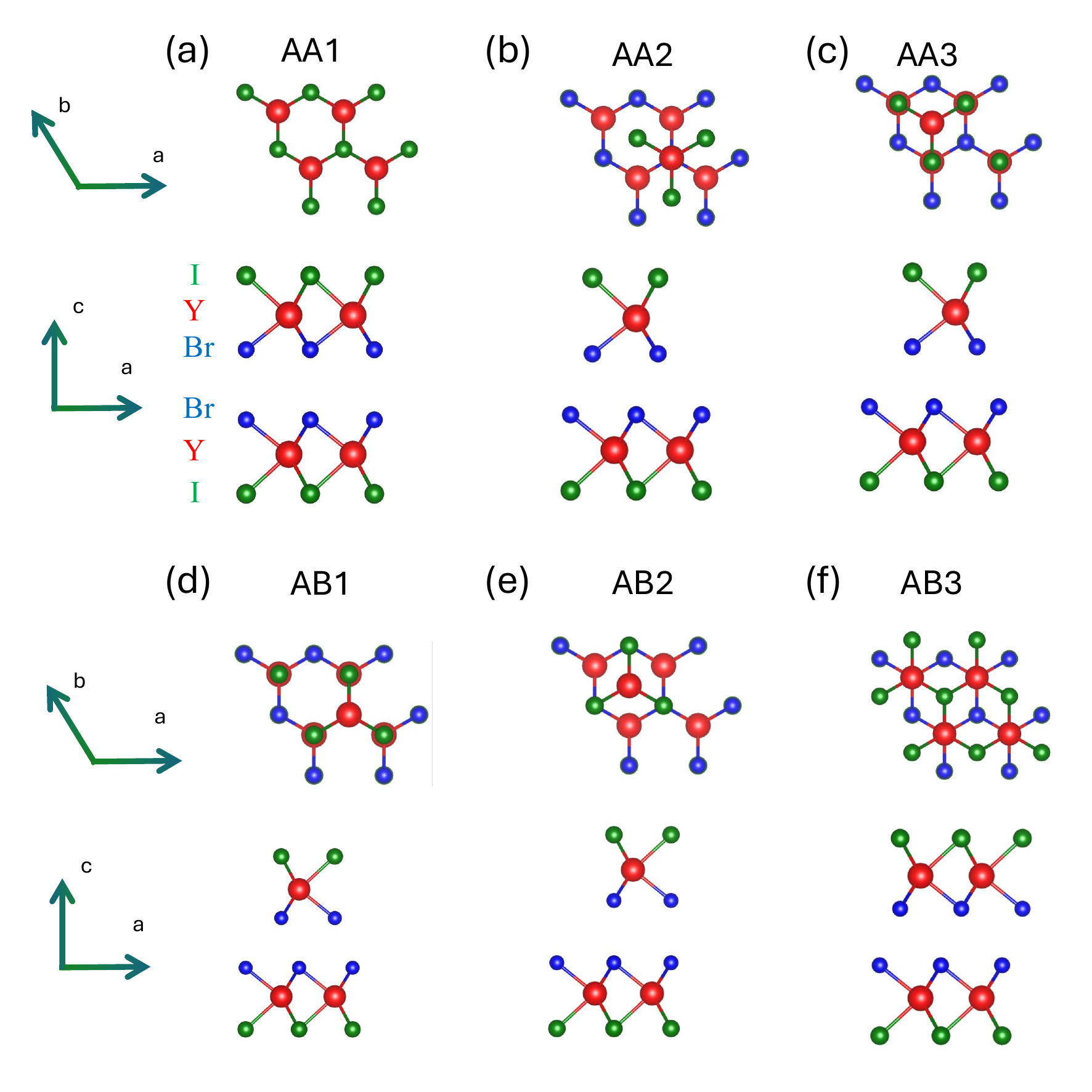}
    \caption{\label{fig0} The top-view and side-view structure of six bilayer Janus YIBr. I,  Y, and Br atoms are green, red, and blue, respectively. The primitive vectors are indicated with the arrows.}
\end{figure}

The bilayer Janus 2D structures have been extensively studied experimentally and theoretically for their strange physical properties\cite{VCl2,YIBr,HTDFT,J01,J02}. Interlayer stacking is a very important degree of freedom for the modulation of physical properties. For the bilayer 2D Janus YIBr(I-Y-Br-Br-Y-I), the I atoms are located at the outermost layer of the bilayer structure and the Br atoms are at the innermost layer of the bilayer structure. There are six different typical stacking configurations with high symmetry, which are shown in Fig. \ref{fig0}. These configurations can be divided into two groups: AA1-AA3(a-c) and AB1-AB3(d-f) according to the symmetry. The AA1 stacking has mirror symmetry perpendicular to the layers and a threefold rotation axis in the plane and AA2 and AA3 can be obtained through a specific interlayer slip. The sliding destroys the mirror symmetry perpendicular to the layers, so the AA2 and AA3 only have a threefold rotation axis in the plane. AB1 is obtained by rotating the upper layer of AA1 by 180°, at which point a twofold rotation axis between the layers appears. AB2 and AB3 can be obtained through a specific interlayer slip and twofold rotation axis still exists after the rotation. All AB configurations have threefold rotation axis in the plane like AA structures.
The lattice constant is $a=$ 3.96\AA{} and the bond length of Y-I and Y-Br  3.12\AA{} and 2.94\AA{}. In order to obtain a relatively reasonable structure, the distance between the two YIBr layers in the initial structure for structural optimization calculation is set smaller than the actual value. The results show the distance between the two monolayers is 3.74 \AA{} for AA1, 3.73 \AA{} for AB2 and 3.19 \AA{} for other configurations. The difference between the two is due to the fact that the Br atoms in the inner layers of the two structures share the same position in the plane, which are shown in Fig. \ref{fig0}(a,e). 

Previous studies show that the magnetism of the YIBr monolayer has ferromagnetic ground state. The monolayer YIBr Janus structure is ferromagnetic with 1 $\mu_B$. In order to explore the magnetic properties of the bilayer structure and find out the direction of the easy axis, we calculated the MAE of the bilayer YIBr of the six interlayer configurations. The results which are shown in Table \ref{tab:AFM_FM_energy} indicate that the AFM configurations have lower energy than the FM configurations so the magnetic ground state of interlayer interaction is AFM for all six configurations. However, the easy axis of the bilayer is different for different configurations. For the AFM AA configurations, the easy axis is out-of-plane along the [001] direction for the AA1 and AA2 configurations and in-plane for the AA3 configurations. And for the AFM AB configurations, the easy axis is out-of-plane along the [001] direction for the AB2 configurations and in-plane for the AB1 and AB3 configurations. The easy axes of the FM are in-plane for the FM magnetic states except for the AB2 configuration. It is acknowledged that the AA structure can be converted into each other by sliding, so the direction of the easy axis can be adjusted easily.

\begin{table*}
    \small
    \caption{Total energies $E$ and energy differences $\Delta E$ for AFM (left) and FM (right) configurations with different easy axes. The [001] direction is out-of-plane, and the [100] and [110] directions are in-plane.}
    \label{tab:AFM_FM_energy}
    \begin{center}
        \renewcommand{\arraystretch}{1.5}
        \begin{ruledtabular}
            \begin{tabular}{cccccccc}
                \multicolumn{4}{c}{\textbf{AFM}} & \multicolumn{4}{c}{\textbf{FM}}                                                                                                                 \\
                \textbf{System}                     & \textbf{direction}              & \textbf{E} (eV)& \textbf{$\Delta$E} (eV)& \textbf{System}         & \textbf{direction} & \textbf{E} (eV)& \textbf{$\Delta$E} (eV)\\
                \colrule
                \multirow{3}{*}{AA1}             & [001]                           & -26.8030   & 0.0000             & \multirow{3}{*}{AA1} & [001]              & -26.7999   & 0.0031             \\
                                                 & [100]                           & -26.8019   & 0.0010             &                      & [100]              & -26.8004   & 0.0026             \\
                                                 & [110]                           & -26.8023   & 0.0007             &                      & [110]              & -26.8004   & 0.0025             \\ \hline
                \multirow{3}{*}{AA2}             & [001]                           & -26.7929   & 0.0101             & \multirow{3}{*}{AA2} & [001]              & -26.7896   & 0.0134             \\
                                                 & [100]                           & -26.7906   & 0.0124             &                      & [100]              & -26.7900   & 0.0129             \\
                                                 & [110]                           & -26.7906   & 0.0124             &                      & [110]              & -26.7900   & 0.0129             \\ \hline
                \multirow{3}{*}{AA3}             & [001]                           & -26.7898   & 0.0132             & \multirow{3}{*}{AA3} & [001]              & -26.7893   & 0.0137             \\
                                                 & [100]                           & -26.7904   & 0.0126             &                      & [100]              & -26.7899   & 0.0131             \\
                                                 & [110]                           & -26.7904   & 0.0126             &                      & [110]              & -26.7899   & 0.0131             \\ \hline
                \multirow{3}{*}{AB1}             & [001]                           & -26.7905   & 0.0125             & \multirow{3}{*}{AB1} & [001]              & -26.7881   & 0.0149             \\
                                                 & [100]                           & -26.7910   & 0.0120             &                      & [100]              & -26.7887   & 0.0143             \\
                                                 & [110]                           & -26.7909   & 0.0121             &                      & [110]              & -26.7886   & 0.0144             \\ \hline
                \multirow{3}{*}{AB2}             & [001]                           & -26.8028   & 0.0002             & \multirow{3}{*}{AB2} & [001]              & -26.8023   & 0.0007             \\
                                                 & [100]                           & -26.8010   & 0.0020             &                      & [100]              & -26.7998   & 0.0032             \\
                                                 & [110]                           & -26.8009   & 0.0021             &                      & [110]              & -26.7999   & 0.0031             \\ \hline
                \multirow{3}{*}{AB3}             & [001]                           & -26.7907   & 0.0123             & \multirow{3}{*}{AB3} & [001]              & -26.7898   & 0.0132             \\
                                                 & [100]                           & -26.7911   & 0.0119             &                      & [100]              & -26.7901   & 0.0129             \\
                                                 & [110]                           & -26.7911   & 0.0119             &                      & [110]              & -26.7901   & 0.0129             \\
            \end{tabular}
        \end{ruledtabular}
    \end{center}
\end{table*}

\subsection{Electronic energy bands}

\begin{figure*}[htbp]%
    \includegraphics[width=\linewidth]{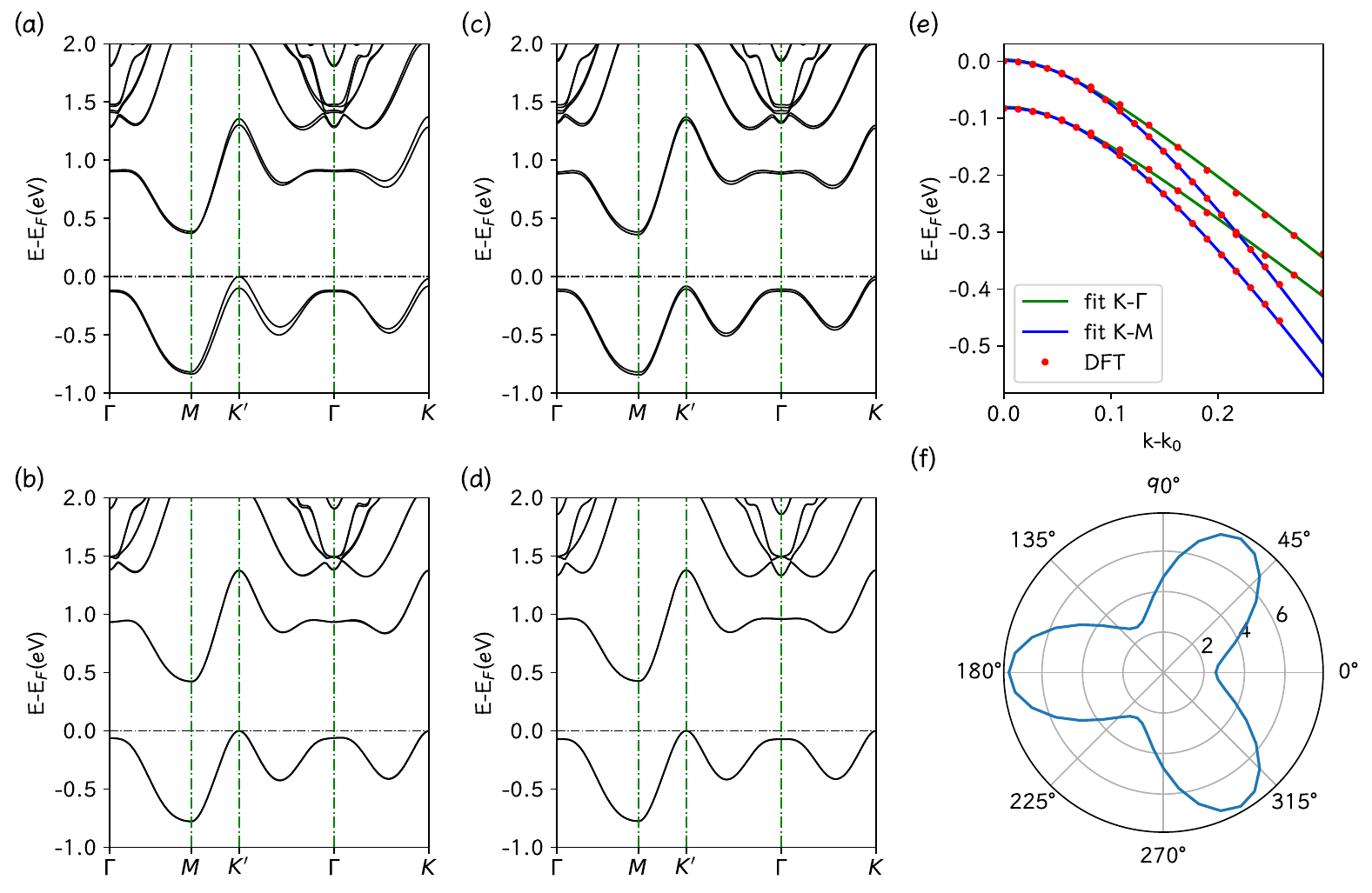}
    \caption{\label{fig2} Band structures for (a) AA1 with an out-of-plane easy axis, (b) AA3 with an in-plane easy axis, (c) AB2 with an out-of-plane easy axis, and (d) AB3 with an in-plane easy axis. (e) Effective model fit (solid lines) to the DFT bands. (f) Angular dependence of $\alpha$ for the AA1 structure with an out-of-plane easy axis.}
\end{figure*}

Interlayer interaction plays a very important role in the layer structure, and further research shows that the interlayer interaction can import SOC from the one layer to another layer. Different stacking configurations with different interlayer interaction may cause the different properties of the electronic structure. So we investigate the electronic band structure for the different stacking configurations. The band structure of AA1 and AB2 with the out-of-plane easy axis are shown in Fig. \ref{fig2} (a,c). The results show that there is spin splitting near the Fermi level in both structures. The band energy at K and K' valleys of the AA1 is roughly equal, but the band structure of AB2 exhibits very obvious valley polarization. The AA3 and AB3 with the in-plane easy axis are shown in Fig. \ref{fig2} (b,d). Unlike the out-of-plane easy axis, the spin  of AA3 and AB3 is degenerate near the Fermi level. The band structure of other stacking configurations and easy axis are shown in Fig. S1 (Supplemental Material). According to the results, there is spin splitting near the Fermi level for the situation of out-of-plane easy axis and there is obvious valley polarization for the situation of AB stacking configuration.

We also find that the energy dispersion relation of the valence band near the Fermi level is close to a Dirac-like relativistic dispersion relation. We use an effective low-energy Hamiltonian model to describe the valence bands and analyse the electronic properties.
\begin{equation}
    H_0(k) =- \sqrt{\alpha k^2 + \beta}
    \label{eq:m1}
\end{equation}
where $\alpha$ and $\beta$ are the parameters from the band fitting. The results of effective model fitting are shown in Fig. \ref{fig2}(e). It is found that this effective model can describe the both valence bands under the Fermi level for all the six stacking configurations with the out-of-plane or in-plane easy axis and the parameters for other structures are shown in Table. S1. The $\alpha$ and $\beta$ of the top valence band and the second valence band are similar. The range of values ​​for $\alpha$ is 2.124 - 2.370 eV$^2 \cdot$\AA{}$^2$ at the line of K-$\Gamma$ and 7.870-8.643 eV$^2 \cdot$\AA{}$^2$ at the line of K-M and $\beta$ is 0.015-0.017 eV$^2$ at the line of K-$\Gamma$ and 0.255-0.294 eV$^2$ at the line of K-M for all six stacking configurations with in- or out-of-plane easy axis.
The results show that the parameters $\alpha$ and $\beta$ have anisotropy and we calculate the parameters for different directions in k space and the results of $\alpha$ of AA1 with out-of-plane easy axis are shown in Fig.\ref{fig2} (f). The anisotropy parameters have 120$^{\circ}$ of symmetry.  The effective mass of the holes in the VBM can be calculated by the effective model $\frac{1}{m^*} = \frac{1}{\hbar^2} \left. \frac{d^2E}{dk^2} \right|_{k=0}$, and the range of effective mass of hole for the AA1 structure with out-of-plane easy axis is 0.41-0.46 m$_e$, where m$_e$ is the free electron mass.

\subsection{Spin, valley and layer coupling}

The results of the calculation of the spin resolved energy bands show that there is obvious spin splitting and valley polarization in the Janus structure. For AA structures, there are 64-98 meV spin splitting at K or K' point when the easy axis is out-of-plane. However, the spin splitting is small, and the energy difference of the two spins is only 3-17 meV when the easy axis is in-plane. For AA structures with an out-of-plane easy axis, the VBM energy difference between K and K′ is at most 17 meV, which is much smaller than the spin splitting. And the energy difference between the K and K' is 0 meV when the easy axis is in-plane for AA structures. For AB structures, the situation is different. The spin splitting becomes smaller and the valley polarization becomes larger. The spin splitting is 17-24 meV when the easy axis is out-of-plane, the spin splitting is 0-1 meV when the easy axis is in-plane. The energy difference between the K and K' is 80-81 meV when the easy axis is out-of-plane, the energy difference between the K and K' is 0 meV when the easy axis is in-plane. So the spin splitting and valley polarization are related to the direction of the easy axis.

\begin{figure}[htbp]%
    \includegraphics[width=\columnwidth]{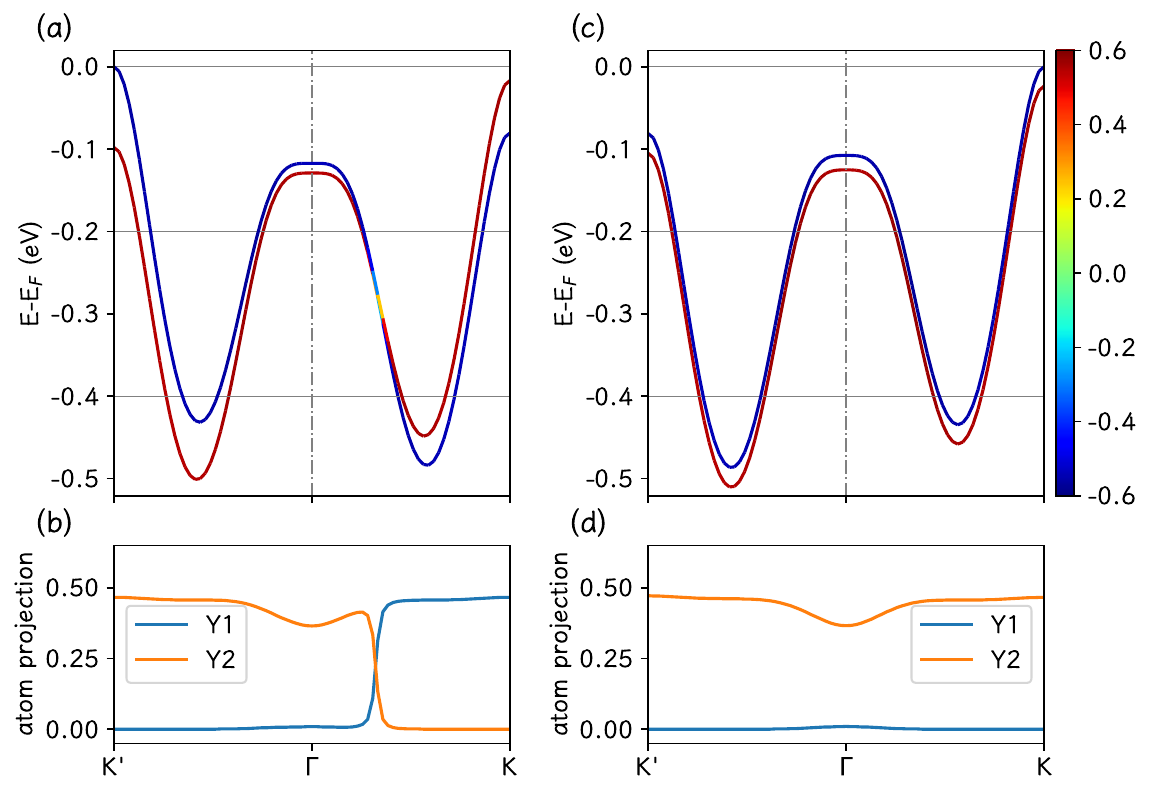}
    \caption{\label{fig3} The spin of the two valence bands for AA1 (a,b) and AB2(c,d) structure and the atoms projection of the top valence band. The Y1 means the Y atom from the lower layer and Y2 means the Y atom from the upper layer.}
\end{figure}

To investigate the relationship between the spin, valley and layers, we calculate the spin and atom projection of the two top valence band along the K-$\Gamma$-K' path. The results of the two typical structures when the easy axis is out-of-plane are shown in Fig. \ref{fig3}. For the AA1 structure, it is clear that the spins of the top valence band are opposite in direction, there is obvious spin splitting at the K and K' and  up spin at K point and down spin at K' point. Moreover, the VBM at K' is only 17 meV higher than that at K. The atom projection weight also shows that the down spin of the VBM at K point is from the upper layer and the up spin of the VBM at K' point is from the lower layer. 

For AB2 structure, the situation is different from the AA structures. The magnitude of spin splitting is smaller than that in the AA2 structure and the band structures of the two valence bands are also similar. The energy difference between the K and K' is larger than that of the AA2 structure. For both K and K' valley, the spin and atom projection weight show that the valence band is spin down and from the upper layer, and the second band under the Fermi level is spin up and from the lower layer.

So we get a uniform effective model which can describe the two valence bands near the Fermi level with different spin and valley according to the coupling of spin, valley and layers.

\begin{equation}
    H(k) = H_0(k) +\lambda_{1}(k)\tau_{z} + \lambda_{2}(k)s_{z} + \lambda_{3}(k)\tau_{z}s_{z}
    \label{eq:m2}
\end{equation}

The H$_0(k)$ is the Dirac Hamiltonian in Eq.\ref{eq:m1}, which describes the massive Dirac-like relativistic dispersion relation. The second term is about the valley polarization and $\lambda_{1}$ describes the valley splitting, $\tau$ = 1 for K valley and -1 for K' valley. $\lambda_{2}$ describes the spin splitting, s$_z$ = 1 for spin up and -1 for spin down. $\lambda_{3}$ describes the spin valley coupling. 
For the AA structures with an out-of-plane easy axis, the genuine valley term is negligible ($\lambda_{1}$ = 0) and the spin split is mainly from the spin valley coupling term($\lambda_{3}$), which makes VBM at the K and K′ valleys have opposite spin. There is also a weak spin term $\lambda_{2}$ which leads to the slight asymmetry of the spin splitting between the two valleys (e.g., 85 vs 78 meV for AA2, which is shown in Table. S1.). This also produces the small VBM energy difference between K and K' (17 meV for AA1, 4 meV for AA2 and 3 meV for AA3 ). For AA2, the $\lambda_{2}$ and $\lambda_{3}$ are -2 meV and 40.8 meV at K or K'. 
And for the AB structures with out-of-plane easy axis, there is obvious valley polarization and the spin of top valence band at K and K' valley are the same, so $\lambda_{3}$ is 0 and $\lambda_{1}$ and $\lambda_{2}$ are not 0. The $\lambda_{1}$ and $\lambda_{2}$ are 40.5  meV and 11 meV at K or K' for AB2 structure. For the in-plane easy axis, the spin and valley are degenerate, which is shown in Fig. S2. So the $\lambda_{1}$ $\lambda_{2}$ and $\lambda_{3}$ are all 0. Moreover the easy axis can be controlled by the sliding of relative position of the two layers. The spin and valley properties are different for the different easy axes, so the sliding can be used to control the spin layer valley coupling and properties indirectly.

\subsection{Berry curvature and valley/layer Hall effects}

\begin{figure}[htbp]%
    \includegraphics[width=\columnwidth]{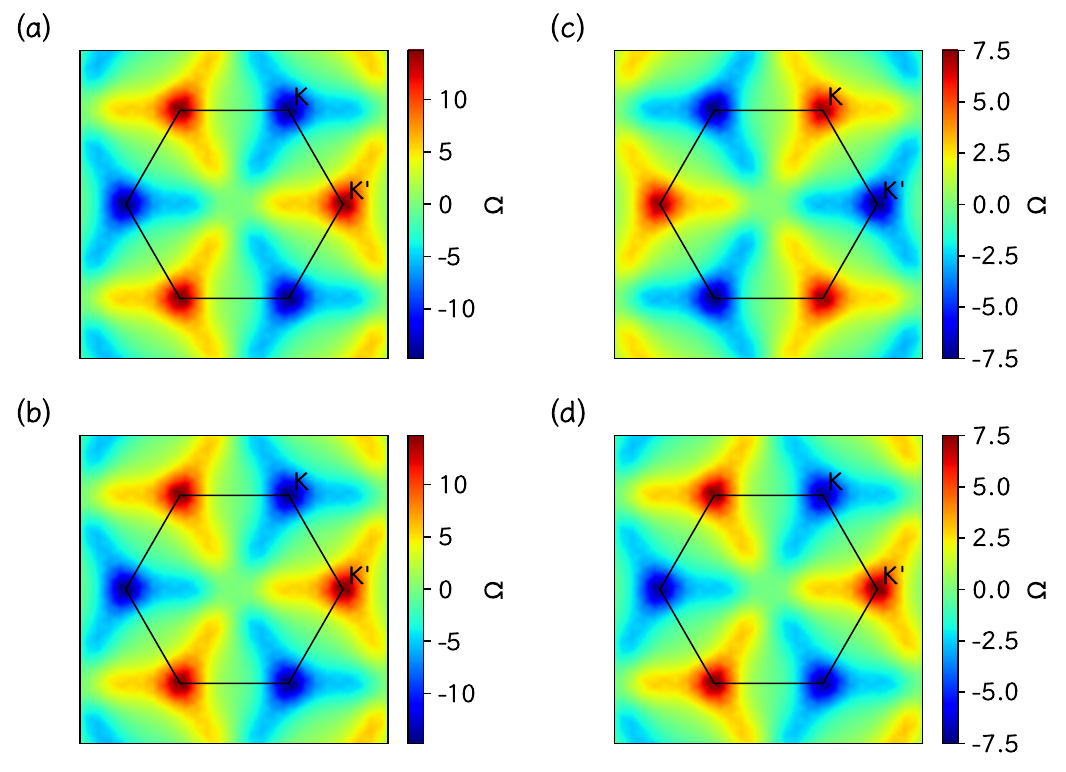}
    \caption{\label{fig4} The summary of the Berry curvature of the two valence bands for AA1 with out-of-plane easy axis(a) and AA3 with in-plane easy axis(b) structure. The Berry curvature of AB2 with out-of-plane easy axis for the upper valence band (c) and the lower valence band (d).}
\end{figure}

To investigate the topological properties of the bilayer structure, we calculate the Berry curvature of the two valence bands in the whole k space. According to Kubo formula\cite{rx47}, the Berry curvature of the $n$-th band at the point $\vec{k}$, $\Omega^n(\vec{k})$, can be written as
\begin{equation}\label{eq:13}
    \Omega^n=i\sum_{m\neq n}\frac{\langle n | \hat{v}_x|m\rangle \langle m |\hat{v}_y|n\rangle - \langle n | \hat{v}_y|m\rangle \langle m | \hat{v}_x|n\rangle}{(E_n-E_{m})^2}
\end{equation}
where $\hat{v}_x=\frac{\partial \hat{H}}{\partial k_x}$ and $\hat{v}_y=\frac{\partial \hat{H}}{\partial k_y}$, $E_n=E_n(\vec{k})$ is the $n$-th band and $|n\rangle=|n\vec{k}\rangle$ (or $\psi_n(\vec{k})$) is the corresponding eigen wavefunction. The results are shown in Fig. \ref{fig4}.
The results show that the configurations within the same group (AA or AB) have similar Berry-curvature distributions. According to the Fig. \ref{fig4}(a,b), AA1 and AA3 have different easy-axis directions, but their Berry curvature distribution for the two valence bands are similar. For each of the two valence bands, the Berry curvature is +7 Å² at K and -7 Å² at K', and the two bands have the same sign at a given valley. As a result, the sum of the two bands is +14 Å² at K and -14 Å² at K', so the integral of the Berry curvature over the entire Brillouin zone vanishes. This valley-contrasting Berry curvature gives rise to a valley Hall effect. Further analysis of the single-band Berry curvature shows that the two bands from different layers carry positive (negative) values at the same K (K') point.

For the AB structures, the Berry curvature is different from that of the AA type, the Berry curvature of two valence bands is shown in Fig. \ref{fig4}(c,d). The upper valence band has Berry curvature +7 Å² at K and −7 Å² at K′, while the lower band has -7 Å² at K and +7 Å² at K'. Thus the two bands from different layers carry opposite signs at the same K (K') point, and the total Berry curvature of the two bands vanishes at each valley. This is the layer-locked hidden Berry curvature, which gives rise to a layer Hall effect\cite{LHE}.

\subsection{External electric field}

\begin{figure}[htbp]%
    \includegraphics[width=\columnwidth]{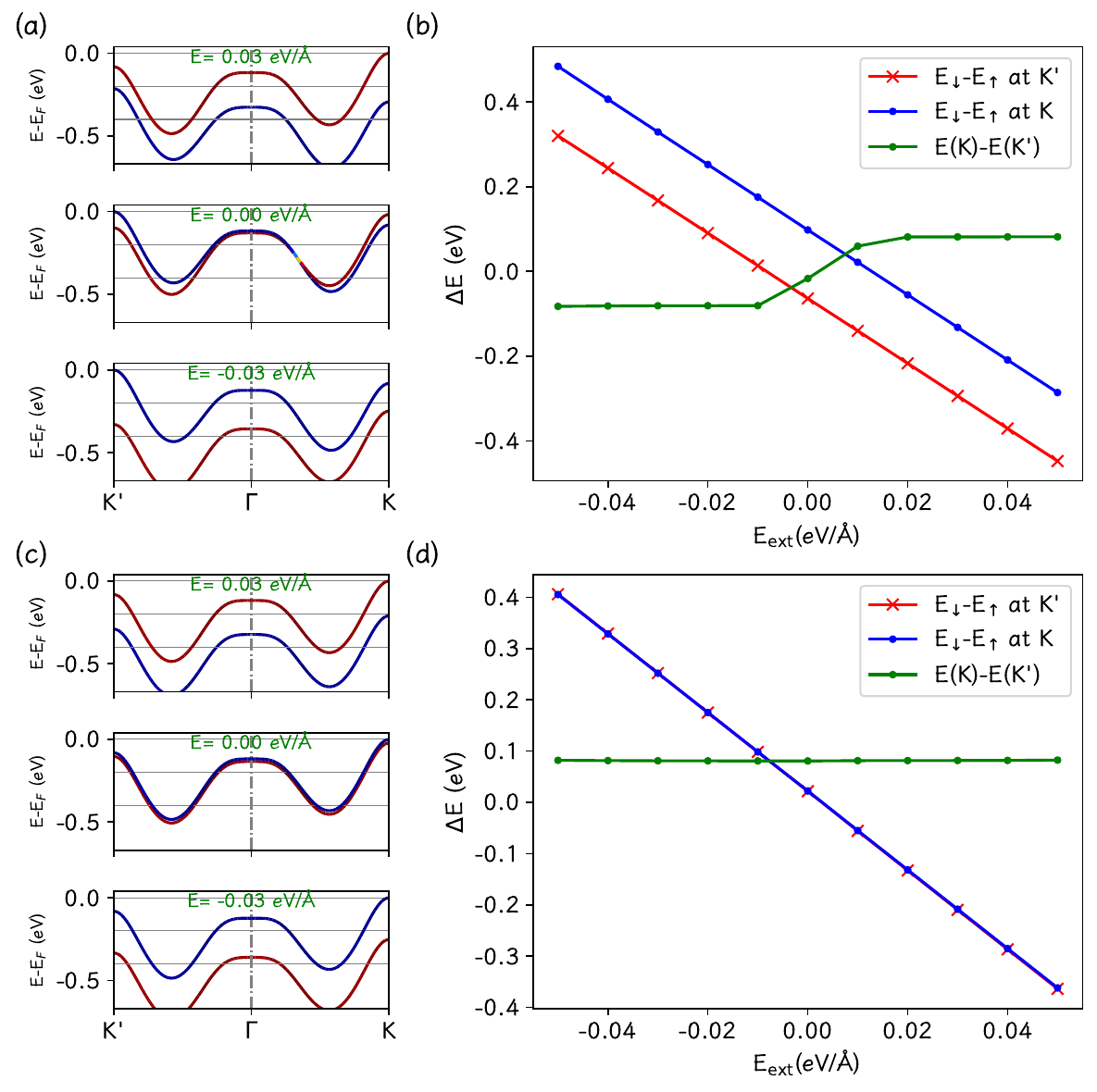}
    \caption{\label{fig5} The energy bands for AA1(a) and AB2(c) under the exernal electric field. The energy differences between the spin up and spin down at K (K') and the valley polarization for AA1(b) and AB2(d).}
\end{figure}

To investigate the effect of external electric fields on electronic properties and possible control methods, we calculate the electronic structure under the out-of-plane external electric field. We apply a -0.05 - 0.05 V/\AA{} external electric field in the z direction and calculate the electronic structure for the six structures. The results of AA1 and AB2 are shown in Fig. \ref{fig5}. For both structures, the electric field can increase the magnitude of spin split, regardless of whether the electric field is upward or downward. So the effective Hamiltonian with external electric field can be written as
\begin{equation}\label{eq:14}
    H_{e}(k) = H(k) + h(E) s_z
\end{equation}
where H$_{e}$ is the Hamiltonian with external electric field, E is the external electric field,  h(E) is a linear function that describes the effect of external electric field on the spin split.

\begin{figure}[htbp]%
    \includegraphics[width=\columnwidth]{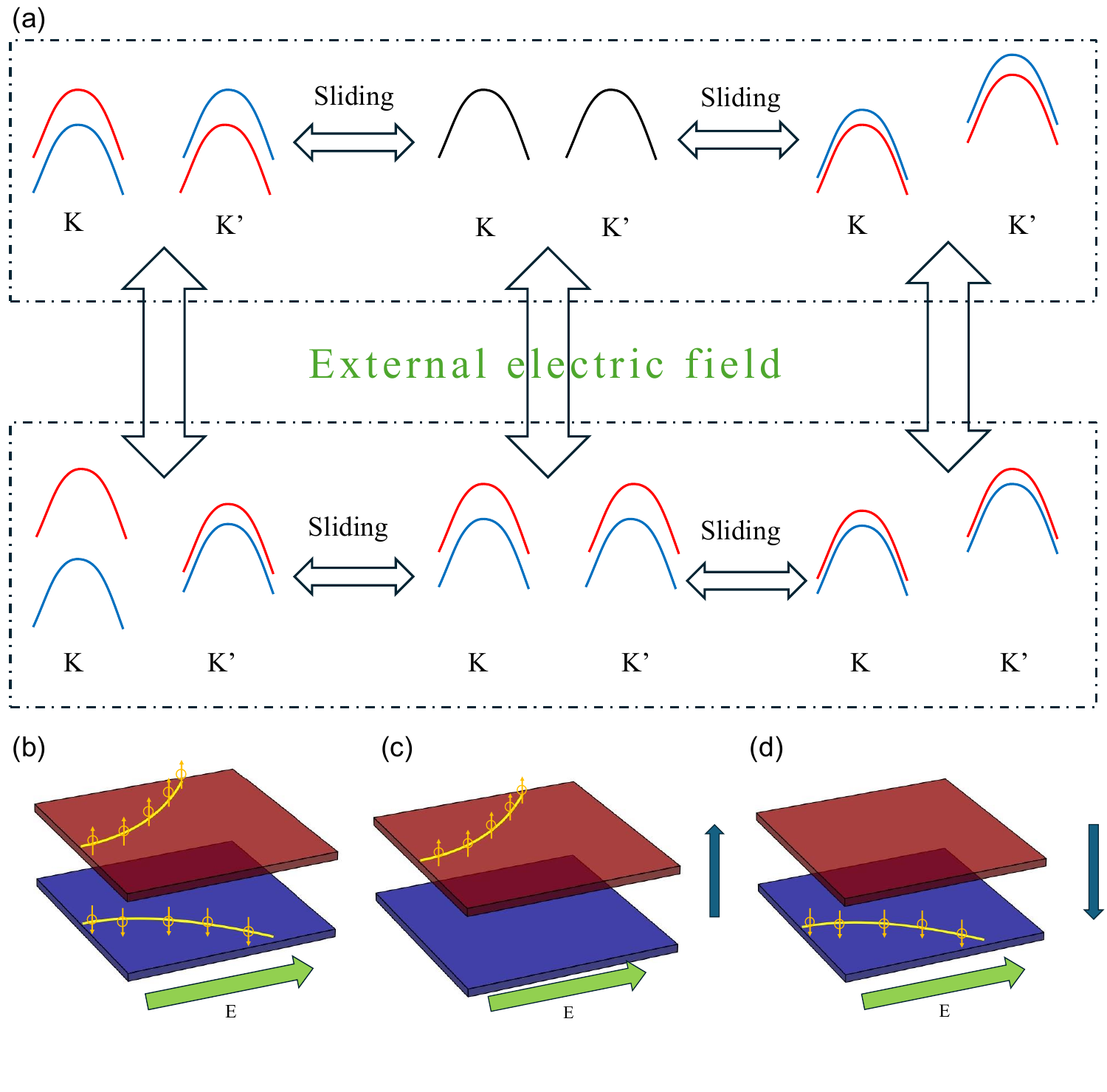}
    \caption{\label{fig6} (a) Joint regulation of electric properties by sliding and external electric field. (b-d) The layer-locked anomalous Hall effects controlled by the electric field. Positive sign (+) means holes.}
\end{figure}

For AA1, the electric field along the z-axis can increase the energy of the band with up spin from the lower layer while an electric field in the opposite direction can increase the energy of the band with down spin from the upper layer. The spin split at K or K' is about 0.4 eV under the 0.05 V/\AA{} external electric field. The external electric field can also influence the valley polarization, which is shown in the green lines of Fig. \ref{fig5}.(b) The alteration of the top valence band can cause the change of the spin, valley and layer of the hole. A valley polarization of 80 meV is achieved at an electric field of 0.02 V/\AA{}, and it can vary between -80 meV and 80 meV in response to changes in the external electric field, which likewise enables a tunable anomalous valley Hall response. For comparison, bilayer YI$_2$ reaches 106 meV at 0.1 V/\AA{}\cite{YI2}, while monolayer MnBr changes by less than 1 meV even at 0.3 V/\AA{}\cite{AFM_MnBr}. The present system thus provides a comparable valley polarization at a smaller field, with much stronger field tunability. The field-controlled spin, valley, and layer states of the holes and the resulting anomalous valley Hall effect are summarized in Fig.\ref{fig6}(b-d). For the AB2 structure, the external electric field can not change the valley polarization, the energy at the K valley is higher than at the K' valley, but the external electric field can change the spin and the layer of the hole. The change of 0.05 V/\AA{} external electric field can cause a change of the spin split of 0.4 eV. This enhanced spin splitting realizes a tunable layer-locked anomalous valley Hall effect with the original valley polarization.

For the condition of in-plane magnetic easy axis, the electric field can also cause the spin splitting. The properties of the spin of the VBM can be controlled by external electric field, but the valley polarization can not be induced by external electric field, and thus the anomalous valley Hall effect is absent in this case, which are shown in Fig. S3 and Fig. S4. For AB structure the magnitude of spin splitting under the same electric field is smaller than that in the out-of-plane easy axis case. 

\subsection{Further discussion}

In this bilayer Janus YIBr system, the different stacking types have different symmetry and the relative position between two layers can have an effect on interlayer interactions, which can lead to different magnetic and electronic properties. Sliding between the two layers can control the easy axis direction and the direction of the easy axis can determine the spin splitting and valley polarization. The out-of-plane easy axis can cause the spin splitting or valley polarization and the spin and valley remain degenerate when the easy axis is in-plane.
The coupling of spin, valley and layer which is determined by the easy axis in the valence band can provide an effective control mechanism for the hole properties. The calculation of the Berry curvature shows that the AA and AB stacking types have different Berry curvature distribution which is related to the symmetry of the stacking type. The external electric field can also modulate the properties of band structure continuously. The spin splitting and valley polarization can be induced and enhanced and the spin, valley and layer of the hole can be controlled by external electric field, as illustrated in Fig.\ref{fig6}(a).

These discoveries can be used as spin/valley filters and multistate memory, the stacking and external electric field can control the hole properties continuously and determine the spin, valley and layer of the hole. The sliding can control the band structure through the direction of the easy axis and external electric field can adjust the magnitude of spin splitting. The 80 meV valley polarization achieved at a low field of 0.02 V/\AA{} is favorable for low-power and low-stress device operation. This combined regulation approach represents an attractive new concept for device design.

\section{CONCLUSION}

In summary, six types of stacking structures of bilayer Janus YIBr are investigated using first-principles calculations. It is found that the stacking configuration can control the easy axis direction. The calculations of band structure show that there is a Dirac relativistic dispersion relation in the valence band in a wide energy window of 0.3eV at least. For the situation that the easy axis is out-of-plane, there is spin splitting or valley polarization for different stacking types and the valley and spin are degenerate when the easy axis is in-plane. The calculation of spin resolved band structure and atom projection shows the relationship between spin, valley, layer and easy axis and we get an effective low-energy Hamiltonian model to describe the energy bands and analyse the electronic properties, including coupling of spin, valley and layer. It is revealed by the Berry curvature calculation that the AA stackings exhibit a valley Hall effect and the AB stackings exhibit a layer Hall effect. We also investigate the external electric field and find that it can induce and enhance the spin splitting, realize an anomalous valley Hall effect, and control the spin, valley and layer of the hole near the Fermi level. Our further analysis reveals the mechanism from the symmetry of the structure to the topological properties and their close relationship and the possible control methods using sliding and external electric field. These can be useful for exploring more properties and functionalities in bilayer Janus structures for promising devices.

\begin{acknowledgments}

The numerical calculations were performed in the Tianhe Xingyi Supercomputer system at the National Supercomputer Center of Guangzhou, Guangzhou, China.

\end{acknowledgments}

\bibliographystyle{apsrev4-2}

\bibliography{b1.bib}

\end{document}